\documentclass[aps,prb,reprint,groupedaddress]{revtex4-1}

\begin{document}


\title{Comment on "Shallow donor states near a semiconductor-insulator-metal interface"}


\author{L.F. Makarenko}
\email{makarenko@bsu.by}
\affiliation{Faculty of Applied Mathematics and Computer Science, Belarusian State University, Independence Ave. 4, 220030 Minsk, Belarus}
\author{O.A. Lavrova}
\affiliation{Faculty of Mechanics and Mathematics, Belarusian State University, Independence Ave. 4, 220030 Minsk, Belarus}


\date{\today}

\begin{abstract}
In a recent paper Hao et al. [Phys. Rev. B 80, 035329 (2009)] reported variational calculations of energy spectrum 
for shallow hydrogenic donor in the structure of semiconductor/insulator/metal with a new type of trial wave function. 
They also performed calculations for semiconductor/insulator system and found that their method gives energy values 
lower than those obtained by MacMillen and Landman [Phys. Rev. B 29, 4524 (1984)]. As follows from these results 
MacMillen and Landman have got much larger errors in energy values than they expected. However we confirm that the 
theoretical approach suggested by MacMillen and Landman gives rather accurate energy values for the system of hydrogenic donor 
near the interface between semiconductor and insulator.
\end{abstract}


\maketitle


In a recent paper \cite{HDAP09}, Hao et al. (HDAP) reported variational calculations of energy spectrum for shallow hydrogenic donor in the structure of semiconductor/insulator/metal with a new type of trial wave function. To evaluate the quality of the suggested trial function HDAP compared the calculated ground state energies at different donor positions with the calculations obtained by MacMillen and Landman \cite{ML84}. The authors of Ref. [\onlinecite{ML84}] ensured that their calculated energy values are accurate to four significant figures. However, in Ref. [\onlinecite{HDAP09}] it was claimed that the errors in the result of Ref. [\onlinecite{ML84}] may be even grater than $9 \%$. The aim of this comment is to resolve this contradiction. We argue that the comparison in Ref. [\onlinecite{HDAP09}] is not adequate.

To calculate the electronic spectrum under consideration an eigenvalue problem for the following equation (written in dimensionless form) should be solved \cite{ML84}:
\begin{equation}\label{Eq}
\left(-\nabla^2 - \frac{2}{r_D}-\frac{2Q^*}{r_{\bar{D}}}+\frac{Q^*}{2z}\right)F(\vec{r})=EF(\vec{r}).
\end{equation}
All distances are scaled to units of effective Bohr radius ($a_B^*$) and the energy $E$ is given in effective Rydbergs ($\mbox{Ry}^*$). The parameter $Q^*$ is given by:
\[
Q^*=\frac{\epsilon_s-\epsilon_{ins}}{\epsilon_s+\epsilon_{ins}},
\]
where $\epsilon_s$ and $\epsilon_{ins}$ are the dielectric constants of the semiconductor and the insulator respectively. The distance of the donor from the interface is $R$ and $r_D$ and $r_{\bar{D}}$ are the distances of an impurity electron from the donor ($D$) and its image ($\bar{D}$), respectively. 

HDAP compared their results calculated for $Q^*=1$ in Eq. (\ref{Eq}) (see Table I in Ref. [\onlinecite{HDAP09}]) to the values calculated in Ref. [\onlinecite{ML84}] for $Q^* \approx 0.8387$ ($\epsilon_s=11.4$, $\epsilon_{ins}=1$). The latter one corresponds to the interface between silicon and vacuum (see Table II in Ref. [\onlinecite{ML84}]). It seems that this incorrectness has been arisen due to the same mistake made in Ref. [\onlinecite{CKS07}].

In order to test this assertion we have performed calculations for $Q^* \approx 0.8387$ using variational and finite-element approaches. For variational calculations a trial wave function was chosen as a sum $\Psi_{trial}=\sum c_{ik} \phi_{ik}$  with the basis wave functions:
\begin{equation}\label{WaveF}
\phi_{ik} = 2 \alpha^{3/2} R \pi^{-1/2}\mbox{exp}\left[-\alpha R (\xi-\eta)\right]L_i(\xi) P_k(\eta),
\end{equation}
where $\xi=\frac{r_D+r_{\bar{D}}}{2R}$ and $\eta=\frac{r_{\bar{D}}-r_D}{2R}$ are prolate spheroidal coordinates, $\alpha$ is a variational parameter, and $L_i(\xi)$ and $P_k(\eta)$ are polynomials of $i$-th and $k$-th degrees, respectively. As $P_k(\eta)$ we choose the Legendre polynomials of odd degree which allow to satisfy the boundary condition $F(\vec{r})=0$ at the interface ($z=0$). Choosing this basis we avoid the necessity of numerical integration. Another advantage of this basis set is a lower number of basis functions which is necessary to obtain the same accuracy for the ground state energy $E_0$ as in Ref. [\onlinecite{ML84}].

To carry out numerical computations using finite element method (FEM), the system is assumed to be rotationally symmetric around $z$-axis and the problem is reformulated in cylindrical coordinates $(r,z)$. The equation is posed in a bounded domain $\Omega=(0,\delta)\times (-\delta-R,0)$ with $\delta=10$  chosen for an approximation of the semi-infinite region $(0,+\infty) \times (-\infty,0)$ with a donor position at $(0,-R)$. Using MATLAB we discretize this problem by linear finite elements on a triangular mesh with a number of unknowns $\approx 150000$. The generalized sparse eigenvalue problem is solved by the implicitly restarted Arnoldi method in MATLAB. 

The comparison of our results for $Q^*\approx 0.8387$ with the theoretical results of MacMillen and Landman is presented in Table \ref{T1} which has the form similar to that one in Ref. [\onlinecite{HDAP09}]. We found that our values are practically coinciding to those obtained by MacMillen and Landman at the same distances from the interface. So our calculations confirm the evaluation of the significant figures for energy values presented by MacMillen and Landman. 

\begin{table}
 \caption{The values of the ground state energy $E_0$ of the impurity electron near the semiconductor-insulator interface for different values of $R/a_B^*$ calculated on the basis of the variational wave functions of  Eq. (\ref{WaveF}) with $Q^*=0.8387$, and $E_0$  found in Ref. [\onlinecite{ML84}] for the same value of $Q^*$.\label{T1}}
 \begin{ruledtabular}
 \begin{tabular}{ccccc}
 $R/a_B^*$&$\alpha$&$E_0/\mbox{Ry}^*$&$E_0/\mbox{Ry}^*$ &$E_0/\mbox{Ry}^*$\\
 &&variational,&FEM,&variational, \\
 &&present&present&Ref. [\onlinecite{ML84}] \\
 \hline \\
0.2&	0.854&	-0.6064&	-0.6062&	-0.6064 \\
0.4&	0.908&	-0.6508&	-0.6507&	-0.6507 \\
0.6&	0.961&	-0.7223&	-0.7222&	-0.7221 \\
0.8&	0.988&	-0.8099&	-0.8098&	-0.8098 \\
1.0&	0.908&	-0.8946&	-0.8946&	-0.8945 \\
1.2&	0.827&	-0.9640&	-0.9643&	-0.9640 \\
1.4&	0.773&	-1.0158&	-1.0164&	-1.0158 \\
1.6&	0.639&	-1.0522&	-1.0530&	-1.0521 \\
1.8&	0.666&	-1.0767&	-1.0771&	-1.0767 \\
2.0&	0.666&	-1.0925&	-1.0932&	-1.0925 \\
3.0&	0.666&	-1.1085&	-1.1089&	-1.1086 \\
4.0&	0.666&	-1.0943&	-1.0952&	-1.0944 \\
5.0&	0.720&	-1.0794&	-1.0804&	-1.0794 \\
6.0&	0.720&	-1.0676&	-1.0679&	-1.0676 \\
 \end{tabular}
 \end{ruledtabular}
 \end{table}

To evaluate the quality of the trial function with two parameters suggested by HDAP we have also performed variational and FEM calculations for $Q^*=1$. The obtained results are presented in Table \ref{T2}. As seen from this table the trial function suggested by HDAP gives rather good bound energy values. However, they are less than the values obtained using the trial wave function consisting of a sum over a basis set of wave functions (as in Ref. [\onlinecite{ML84}] or ours) or by FEM.

\begin{table}
 \caption{The values of the ground state energy $E_0$ of the impurity electron near the semiconductor-insulator interface for different values of $R/a_B^*$ calculated on the basis of the variational wave functions of  Eq. (\ref{WaveF}) with $Q^*=1$, and $E_0$ found in Ref. [\onlinecite{HDAP09}] for the same value of $Q^*$.\label{T2}}
 \begin{ruledtabular}
 \begin{tabular}{ccccc}
$R/a_B^*$&$E_0/\mbox{Ry}^*$&$E_0/\mbox{Ry}^*$ &$E_0/\mbox{Ry}^*$ & Realtive\\
 &variational,&FEM,&variational,& error\\
 &present&present&Ref. [\onlinecite{HDAP09}]& in \%\\
 \hline \\0.4&	-0.7208&	-0.7208&	-0.716&	0.6\\
1.0&	-0.9491&	-0.9491&	-0.927&	2.3\\
1.6&	-1.0917&	-1.0925&	-1.077&	1.4\\
2.0&	-1.1256&	-1.1264&	-1.116&	0.9\\
3.0&	-1.1323&	-1.1327&	-1.128&	0.4\\
4.0&	-1.1131&	-1.1139&	-1.111&	0.3\\
6.0&	-1.0806&	-1.0810&	-1.080&	$<0.1$\\
 \end{tabular}
 \end{ruledtabular}
 \end{table}

HDAP also examined the quality of their trial functions with two and three parameters by comparison with results of their FEM calculations. As it was found in Ref. [\onlinecite{HDAP09}] the use of three parameter functions leads to a lower error. It was estimated as about $1 \%$ when the distance from semiconductor/insulator interface is equal to $a_B^*$. 

However, as it was shown in Ref. [\onlinecite{HDAP09}], the errors in energy values calculated even with the three parameter function become much greater in the case $Q^*=-1$ (semiconductor/metal interface). These errors grow when the donor location tends to the interface. 
The ground energy values were calculated in Ref. [\onlinecite{HDAP09}] by FEM, with three-parameter and two-parameter trial functions. 
They were determined for $R = a_B^*$  as equal to $E_0/\mbox{Ry}^*=0.304, 0.284, 0.276$, respectively. At the same time the approach used in Ref. \onlinecite{ML84} gives the value of $E_0/\mbox{Ry}^*=0.3048$ (with the use of 12 functions in our $\Psi_{trial}$) which is rather close to our FEM value $E_0/\mbox{Ry}^*=0.3050$. 

For the system under consideration a simple single-parametric function has been suggested earlier in Ref. [\onlinecite{SYYS93}]. This function allows to calculate all energies analytically, but provide less accurate energy values as compared with the results in Ref. [\onlinecite{HDAP09}]. 
Another type of a two-parametric trial function has been recently suggested in Ref. [\onlinecite{Ma09}]. This function allows to perform main calculations analytically and gives relative errors for energy values less than $1.4 \%$ for any donor distances from the interface. However all these trial functions do not provide the 'spectroscopic accuracy' of calculated energies.

In conclusion, we confirm that the theoretical approach suggested by MacMillen and Landman \cite{ML84} gives rather accurate energy values for the system of hydrogenic donor near the interface between semiconductor and insulator. One can use their results as a reference in searching for other types of trial functions considering electronic properties of such systems. 

\bibliography{Comment-PRB-Makarenko-2010}

\end{document}